\documentstyle[12pt,epsf,epsfig]{article}
\def\gappeq{\mathrel{\rlap {\raise.5ex\hbox{$>$}}
{\lower.5ex\hbox{$\sim$}}}}
\def\permil{$\%\raise.20ex\hbox{$_0$}}
\def\lappeq{\mathrel{\rlap{\raise.5ex\hbox{$<$}}
{\lower.5ex\hbox{$\sim$}}}}
\begin{document}
\topmargin -1.0cm
\oddsidemargin -0.8cm
\evensidemargin -0.8cm
\pagestyle{empty}
\begin{flushright}
{
hep-th/9806109 \\
SCIPP-98/23 \\
}
\end{flushright}
\vspace*{10mm}
\begin{center}
{\Large\bf Comment on Multigraviton Scattering}\\
{\Large\bf in the Matrix Model}\\
\vspace{2cm}
{\large\bf Robert Echols and Joshua P. Gray}\\
\vspace{.7cm}
{Santa Cruz Institute for Particle Physics}\\
{University of California, Santa Cruz, CA 95064}\\
{echols, jgray@physics.ucsc.edu}\\
\end{center}
\vspace{2cm}
\begin{abstract}
We show by explicit calculation 
that the matrix model effective action does not contain the term
$v_{12}^{2}v_{23}^{2}v_{13}^{2}/R^{7}r^{7}$,
in the limit $R \gg r$,
contradicting a result reported recently.
\end{abstract}
\vfill
%\begin{flushleft}
%January 1998
%\end{flushleft}
\eject
\pagestyle{empty}
%\clearpage\mbox{}\clearpage
\setcounter{page}{1}
\setcounter{footnote}{0}
\pagestyle{plain}
 
%%%%%%%%%%%%%%%%%%%%%%%%%%%%%%%%

\section{Introduction}

The conjectures of \cite{bfss} and \cite{sus} along with the
arguments provided by \cite{sei,sen}, 
and numerous other pieces of evidence (for reviews, see \cite{rev}
and references therein) give one reason to believe that finite $N$
matrix theory describes the discrete light-cone quantization 
(DLCQ) of M-theory with DLCQ supergravity as its low energy limit.  
%There are still potential
%subtleties in regards to whether the finite $N$ matrix model should
%describe DLCQ supergravity \cite{pol}, but the work of \cite{bil}
%seems to indicate that M-theory does have a well defined light-like
%limit.
Although there are still many open questions in the matrix
formulation of M-theory, we would like to focus on whether the three graviton
scattering calculation of \cite{dr} shows a discrepancy between the
matrix model and supergravity.\footnote{See note added for the resolution
to this discrepancy.}

It has recently been reported in \cite{fer}
that supergravity and the matrix model do not disagree on multi-graviton
scattering.
In this note we will show that the term computed in \cite{fer} does
indeed have a supersymmetric cancellation and that the matrix model
effective action does not contain a term of the form 
$v_{12}^{2}v_{23}^{2}v_{13}^{2}/R^{7}r^{7}$.
%\footnote{It has been
%shown in \cite{us2} that the matrix model {\it $S$-matrix} does
%contain this term in agreement with the supergravity $S$-matrix
%calculation in \cite{dr}}.  
%This leaves open the
%possibility that something else is wrong with the matrix model
%calculation or perhaps that there is a problem with the previous
%interpretation of DLCQ supergravity.

It is worthwhile
to review the problems which arise when
one tries to compare three graviton scattering in the matrix model picture
with supergravity, setting the stage for our notation which will be
used in this note. Briefly, the authors of \cite{dr} considered the case of  
three gravitons; two separated a distance $r$ from each other and another a 
distance $R$ from the other two in the limit $R \gg r$. A term in the 
supergravity 
$S$-matrix for three graviton scattering in the small momentum transfer limit
was shown to be
\begin{equation}
\frac{(k_{1}\cdot k_{2})(k_{1}\cdot k_{3})(k_{2}\cdot k_{3})}{q_{1}^{2}q_{2}^{2}}
\end{equation}
where $k_{i}$ are the ith graviton momenta and $q_{1,2}$ are the two relevant momenta transfer. In the language of matrix theory, this corresponds to taking the Fourier transform of the two-loop effective potential
\begin{equation}
\frac{v_{12}^{2}v_{13}^{2}v_{23}^{2}}{R^{7}r^{7}}
\end{equation}
where $v_{12}=(v_{1}-v_{2})$, etc. refer to the relative velocities of the $D0$-branes. The two scales $R$ and $r$ arise from integrating out the  massive degrees of freedom
introduced by giving the diagonal generators
of $SU(3)$ vacuum expectation values:
\begin{equation}
<X^{a}_{i}>=r\delta^{a3}\delta_{i1}+R\delta^{a8}\delta_{i2}
\end{equation}
where $X^{a}_{i}$ are the $9$ $SU(3)$-valued fields describing the bosonic coordinates.
Since $\dot{X_{i}}=\dot{X_{i}^{8}}T^{8}+\dot{X_{i}^{3}}T^{3}$, one
can work out $v_{12}^{2}$, etc. in terms of $\dot{X}^{3}_{i}$
and $\dot{X}^{8}_{i}$
\begin{equation}
v_{23}^{2}\sim(\dot{X}^{3}_{i})^{2}
\end{equation}
\begin{equation}
v_{13}^{2}\sim(\dot{X}^{3}_{i})^{2}+(3\dot{X}^{8}_{i})^{2}-6\dot{X}^{8}_{i}\dot{X}^{3}_{i}
\end{equation}
\begin{equation}
v_{12}^{2}\sim(\dot{X}^{3}_{i})^{2}+(3\dot{X}^{8}_{i})^{2}+6\dot{X}^{8}_{i}\dot{X}^{3}_{i}
\end{equation}
Multiplying these three together yields the expected result for matrix theory
\begin{equation}
v_{12}^{2}v_{13}^{2}v_{23}^{2}\sim (\dot{X}^{3}_{i})^{2}(\dot{X}^{8}_{i})^{4}+(\dot{X}^{3}_{i})^{6}+(\dot{X}^{3}_{i})^{4}(\dot{X}^{8}_{i})^{2}-(\dot{X}^{3}_{i}\dot{X}^{8}_{i})^{2}(\dot{X}^{3}_{i})^{2}.
\end{equation}
In \cite{dr} it was argued that matrix theory was incapable of
reproducing the term,
\begin{equation}
 \frac{(\dot{X}^{8}_{i})^{4}(\dot{X}^{3}_{i})^{2}}{R^{7}r^{7}}
\label{2loop}
\end{equation}
with the
correct powers of $R$ and $r$ at two-loops. In \cite{fer}, it was argued that this term can arise at two-loops from vertices with three massive bosons in the form of the setting-sun diagram, as well as from other two-loop interactions. After describing the background field method used in this note, we go on to show that the one-loop effective operator needed to arrive at the conclusion of \cite{fer} does indeed cancel among bosons and fermions. By exploiting the fact that $\dot{X}_{i}^{8}$ only couples to fields of scale $R$, we
integrate out these most massive modes to find that the first term
containing coupling between the heavy and light states
without supersymmetric cancellations has the form
$(\dot{X}_{i}^{8})^{4}({X}_{i}^{a})^{2}/R^{9}$ as described in \cite{dr}.
Then integrating over the light $SU(2)$ modes of scale $r$ (a=1,2),
we demonstrate that the term in the matrix model 
effective action with four powers of $\dot{X}_i^8$ and the least suppression
in $R$ is
$(\dot{X}_{i}^{8})^{4}(\dot{X}_{i}^{3})^{2}/R^{9}r^{5}$. 

\section{Contributions to the low energy effective action}

The matrix model Lagrangian is obtained from the dimensional reduction of ${\cal{N}}=1$
supersymmetric Yang-Mills theory in $D=9+1$ down to $D=0+1$ dimensions \cite{bfss}.
For our purposes it will be useful to initially keep the action in its ten dimensional
form expressed as

\begin{equation}
S=\int d^{10}x \left(-\frac{1}{4g} F_{\mu \nu}^a F^{\mu \nu a} + \frac{i}{2}
 \overline{\Psi}^a
\Gamma^{\mu} D_{\mu} \Psi^a \right)
\end{equation}
where the field strength is given by
\begin{equation}
F_{\mu \nu}^a = \partial_{\mu} A_{\nu}^a - \partial_{\nu} A_{\mu}^a + f^{abc} A_{\mu}^{b} 
A_{\nu}^{c},
\end{equation}
and the $32 \times 32$ dimensional Dirac matrices $\Gamma$ satisfy the usual algebra 
\(\left\{ \Gamma_{\mu} , \Gamma_{\nu} \right\} = 2 g_{\mu \nu}\) with
metric $g_{\mu \nu}=$diag$(+1,-1,...,-1)$.   
The $32$ component Majorana-Weyl adjoint spinor
$\Psi^a$ has only $16$ real physical components off mass shell.  
%Find a new place to put this!?
We should mention that the center of
mass motion of the $D0$ particles has been removed and we 
will be considering the $SU(3)$ theory with the gauge index a=1-8.

To calculate the one loop contributions to the effective action, we will use the
background field method \cite{ps} and break the gauge field up into
a classical background field and a fluctuating quantum field,
\begin{equation}
A_{\mu}^a \to X_{\mu}^a + A_{\mu}^{'a}, 
\end{equation}
and choose our gauge fixing condition,
$D^{\mu}A_{\mu}^{'a}=0$,   
to be covariant with respect to the background field, 
$D_{\mu}= \partial_{\mu}-it^a X_{\mu}^a$.
%Using the generators for Lorentz transformations on 4-vectors,
By only keeping terms quadratic in the quantum fields, one obtains the gauge-fixed
Lagrangian in the Feynman-`t Hooft gauge:
\begin{equation}
{\cal{L}} = {\cal{L}}_B+{\cal{L}}_{A'}+{\cal{L}}_{\psi}+{\cal{L}}_c.
\end{equation}
The first piece of the Lagrangian just contains the background gauge field,
\begin{equation}
{\cal{L}}_B=-\frac{1}{4g} F_{\mu \nu}^a F^{\mu \nu a}
\end{equation}
whereas the other pieces are quadratic in their respective quantum fields and
contain the background gauge field in the background covariant derivative squared,
$D^2$, as well as in the background field strength $F_{\rho \sigma}^b$:
\begin{equation}
{\cal{L}}_{A'}=-\frac{1}{2g} \left\{A_\mu^{'a}\left[ -\left(D^2 \right)^{ac} g^{\mu \nu}
+\left(F_{\rho \sigma}^b {\cal{J}}^{\rho \sigma} \right)^{\mu \nu} 
\left(t^b\right)^{ac}\right] A_\nu^{'c} \right\}
\end{equation}
\begin{equation}
{\cal{L}}_{\psi}=\frac{1}{2}\,\overline{\Psi}^a \left[\sqrt{ -\left(D^2 \right)^{ac}
+\left(F_{\rho \sigma}^b S^{\rho \sigma} \right) 
\left(t^b\right)^{ac} } \right] \Psi^c
\end{equation}
\begin{equation}
{\cal{L}}_c=\overline{c}^a\left[ -\left(D^2 \right)^{ab}\right] c^b
\end{equation}
where
\begin{equation}
\left({\cal{J}}^{\rho \sigma} \right)_{\alpha \beta} = i\left(\delta_\alpha^\rho
\delta_\beta^\sigma - \delta_\alpha^\sigma \delta_\beta^\rho \right)
\end{equation}
\begin{equation}
S^{\mu \nu}= \frac{i}{4} \left[\Gamma^\mu, \Gamma^\nu \right].
\end{equation}
The one loop effective action is obtained by evaluating the functional
integral for the quantum fields,
\begin{equation}
e^{i \Gamma [X]}=\int {\cal{D}}A'{\cal{D}}\overline\Psi
{\cal{D}}\Psi  {\cal{D}}\overline c {\cal{D}}c \, \exp[i\int d^{10}x 
({\cal{L}}_B+{\cal{L}}_{A'}+{\cal{L}}_{\psi}+{\cal{L}}_c)],
\label{path}
\end{equation}
giving
\begin{eqnarray}
\Gamma[X]=\int d^{10}x (-\frac{1}{4g} F_{\mu \nu}^a F^{\mu \nu a}) 
+ \frac{i}{2} \ln Det[ -\left(D^2 \right) g^{\mu \nu}
+\left(F_{\rho \sigma}^b {\cal{J}}^{\rho \sigma} \right)^{\mu \nu} 
t^b] \nonumber \\ 
- \frac{i}{8} \ln Det[ -\left(D^2 \right)
+\left(F_{\rho \sigma}^b S^{\rho \sigma} \right) 
t^b ] - i \ln Det[ -(D^2)].
\label{det}
\end{eqnarray}
For the fermion functional integration the extra factor
$\frac{1}{4}$ arises from the fermion field having $16$ real components
instead of $32$ complex ones.

To compute the determinants for the different fields, it is useful
to expand
$D^2$,
\begin{equation}
-D^2=-\partial^2+\triangle_1 +\triangle_2
\label{d2}
\end{equation}
where
\begin{equation}
\triangle_1=it^a(\partial_\mu X^{\mu a} + X_\mu ^a \partial^\mu)
\end{equation}
\begin{equation}
\triangle_2=X_\mu^a t^a X^{\mu b} t^b.
\end{equation}
At this point it is convenient to dimensionally reduce to 1-D while 
choosing $X_0^a=0$, so $\triangle_1=0$.  By letting
$X_\mu^a \to r \delta_3^a \delta_\mu^1+ R \delta_8^a \delta_\mu^2
+X_\mu^a$ we can break 
$SU(3) \to U(1) \times U(1)$ giving 
\begin{equation}
\triangle_2=-r^2 t^3 t^3 - 2r X_1^a t^a t^3-R^2 t^8 t^8 - 2R X_2^a t^a t^8
-X_i^a t^a X^{i b} t^b
\end{equation}
with the Latin index going 1--9 and fields $X_i^a$
depending only on time.
It is important to note that in 1-D, r and R are dynamical variables
and we are holding them fixed in the spirit of doing a Born-Oppenheimer
approximation.
The magnetic moment interaction for the bosons
\begin{equation}
\triangle_J^B=\left(F_{\rho \sigma}^b
 {\cal{J}}^{\rho \sigma} \right)^{\mu \nu}t^b
\end{equation}
dimensionally reduced becomes
\begin{equation}
\triangle_J^B=2\left(\partial_0 X_i^b
 {\cal{J}}^{0i} \right)^{\mu \nu}t^b
\end{equation}
since we will be working in a flat direction.
Similarly for the fermions one has
\begin{equation}
\triangle_J^\psi=2\left(\partial_0 X_i^b
 S^{0i} \right)t^b.
\end{equation}
The general form of a determinant in (\ref{det}) can be written
\begin{equation}
Tr \ln(-\partial_0^2+\triangle_2 +\triangle_J).
\end{equation}
Because we are interested in the limit $R \gg r$ and will be letting
only the most massive modes (scale $R$) run in the loop 
(gauge index a=4-7) then
$t^3 t^3 r^2 = \frac{1}{4} r^2$ and $ t^8 t^8 R^2 = \frac{3}{4} R^2$.
It is convenient to rescale, $r \to 2r$, $R \to \frac{2}{\sqrt{3}} R$
and define
\begin{equation}
\triangle_F=\frac{1} {-\partial_0^2-R^2-r^2}
\end{equation}
in addition to
\begin{equation}
\triangle_r= -4r X_1^a t^a t^3
\end{equation}\begin{equation}
\triangle_R= -\frac{4}{\sqrt{3}}R X_2^a t^a t^8
\end{equation}
\begin{equation}
\triangle_2'= -X_i^a t^a X^{i b} t^b
\end{equation}
then the trace becomes
\begin{equation}
Tr\ln(-\partial_0^2-R^2-r^2) + Tr\ln [1+\triangle_F(\triangle_2'
+\triangle_r+\triangle_R+\triangle_J)].
\label{lnn}
\end{equation}
The first piece involving $-\partial_0^2-R^2-r^2$ is a constant and the second
contains the one loop quantum corrections to the effective action which
we will evaluate below by
expanding the logarithm for various numbers of external background fields.
We will find that the first non-zero terms contain four derivatives
even if one just integrates over the most massive modes, $R$.

\subsection{Terms with no derivatives}

We will display in this section a supersymmetric
cancellation between bosons and fermions for all operators
which can be constructed from $-D^2$. 
Even before considering the expansion of $-D^2$ in (\ref{d2}),
it is straightforward to see that all terms in the one loop effective action
with no derivatives cancel.  This is because the
determinants of the bosons and fermions
differ only by derivative terms, and there are an
equal number of bosonic and fermionic
factors in the determinant.  Given that a non-derivative
operator is particularly important in the analysis of \cite{fer},
we show explicitly in this section how non-derivative
operators are cancelled.

The operator in question has the form
%Crucial to the argument in \cite{fer} is the existence of a one loop
%velocity independent local effective operator of the form

\begin{equation}
\delta {\cal{L}}_{eff}^{(1)}= \frac{r^2}{R^3} x_1^b x_1^b
\label{cru}
\end{equation}
where the gauge index, b=1-2, for the small mass $SU(2)$ subgroup (scale r).
Such a term arises from expanding the logarithm in (\ref{lnn})
and is given by
\begin{equation}
-\frac{1}{2} Tr\left[ \triangle_F \triangle_r \triangle_F \triangle_r \right]
\label{2ptr}
\end{equation}
or in frequency space
\begin{equation}
-8Tr[t^3t^at^3t^b]r^{2} \int \frac{dw_1}{2\pi} x_1^a(w_1)
x_1^b(-w_1) \int \frac{dw}{2\pi} \frac{1}{(w^2-R^2)}\frac{1}{[(w-w_1)^2-R^2]}
\label{cru2}
\end{equation}
where we have dropped $r^{2}$ in $\triangle_{F}$ for the leading $1/R$ behavior. Integrating (\ref{cru2}) in the limit $w_1 \to 0$ and then Fourier
transforming gives (\ref{cru}).
Now the important point to notice is that $\triangle_r$ arises
from $-D^2$ which occurs in each determinant for the gauge, fermion and 
ghost fields (\ref{det}).  However, they each give a different
contribution to 
$Tr[t^3t^{a}t^3t^{b}]\sim \delta^{a b}d(j)$, where $d(j)$ is the number of 
components for the various fields
\begin{eqnarray}
d(j)^\psi=32 \; \; \; \; \; \; \; d(j)^{A'}=10 \; \; \; \; \; \; \; d(j)^c=1.
\end{eqnarray}
Now it becomes clear that all terms coming from $-D^2$ in each of the three 
determinants appearing in (\ref{det}) will cancel. To be explicit
one gets
\begin{equation}
[\frac{i}{2}(10)-\frac{i}{8}(32)-i(1)][ \frac{r^2}{R^3} x_1^b x_1^b ]=0.
\end{equation}
A similar result holds for any number of external fields without
derivatives involving $\triangle_r$, $\triangle_R$, $\triangle_2'$.

\subsection{Cancellation of $(F_{0i}^a)^2$ or $(\dot{X}_i^a)^2$}

In this section, we show that terms with two derivatives
cancel as well.  This result is familiar in higher dimensions,
where it is well known that the kinetic terms
of the fields are not renormalized.

Based on the arguments given above the only possible non-vanishing
term with two external fields contains two derivatives and is given by
\begin{equation}
-\frac{1}{2} Tr\left[ \triangle_F \triangle_J \triangle_F \triangle_J \right].
\label{2ptj}
\end{equation}
The supersymmetric cancellation of (\ref{2ptj}) 
between bosons and fermions requires the determination
of $Tr[S^{0i}S^{0j}]=8g^{ij}$ for fermions and 
$Tr[{\cal{J}}^{0i}{\cal{J}}^{0j}]=2g^{ij}$ for bosons. Putting the term
into (\ref{det}) gives
\begin{equation}
[\frac{i}{2}(2)-\frac{i}{8}(8)-i(0)]Tr[t^a t^b]
 \int \frac{dw_1}{2\pi}w_1^2 X_i^a(w_1)
X^{ib}(-w_1) \int \frac{dw}{2\pi} \frac{1}{(w^2-R^2)}
 \frac{1}{[(w-w_1)^2-R^2]}=0
\label{v2can}
\end{equation}
which shows that the 2-point contribution to the effective 
action at one-loop is zero. We can also generalize this result to show that
all possible non-derivative insertions on a loop with two
derivatives will not give a contribution to the effective action.

\subsection{$V^4/R^7$}

Since all terms with two derivatives, no derivatives, or
a mixture cancel
by the arguments given above, the only possible non-vanishing 
term with four external
fields is the four derivative term given by
%\begin{equation}
%-\frac{1}{4} Tr[\triangle_F\triangle_J\triangle_F\triangle_J
%\triangle_F\triangle_J\triangle_F\triangle_J]
%\end{equation}
\begin{equation}
-\frac{1}{4} Tr[({\triangle_F\triangle_J})^4]
\end{equation}
or in frequency space
\begin{equation}
 \int 
\frac{dw_2 dw_3 dw_4 dw}{{(2\pi)}^4} \frac{-(w_2+w_3+w_4) X_i^8[-(w_2+w_3+w_4)]
w_2 X_j^8(w_2) w_3 X_k^8(w_3) w_4 X_l^8(w_4)}
{ [(w+w_2)^2-R^2][(w+w_2+w_3)^2-R^2][(w+w_2+w_3+w_4)^2-R^2][w^2-R^2]}
\end{equation}
with the prefactor
\begin{equation}
-\frac{1}{4} \, 2^4 \, Tr[{(t^8)}^4] \, Tr[{( {\cal{J}}^{0i} )}^4]
\label{tr}
\end{equation}
for the gauge boson  case.  An identical result holds for fermions if one
replaces the Lorentz generator trace with
\begin{equation}
Tr[ S^{0i}S^{0j}S^{0k}S^{0l}]=2(g^{ij}g^{kl}-g^{ik}g^{jl}+g^{il}g^{jk})
\end{equation}
whereas for the gauge bosons one finds
\begin{equation}
Tr[{\cal{J}}^{0i} {\cal{J}}^{0j} {\cal{J}}^{0k} {\cal{J}}^{0l}]=
(g^{ij}g^{kl}+g^{il}g^{jk}).
\end{equation}
Now using (\ref{det}) and the low energy approximation
$w_1,w_2,w_3,w_4 \to 0$, we get
\begin{equation}
-\frac{27i}{4}[(F_{0i}^8)^2]^2 \int \frac{dw}{2\pi} \frac{1}{(w^2-R^2)^4}.
\end{equation}
The integral can be performed in the complex plane using the usual 
$+i\epsilon$ prescription for handling the poles.
Defining $(\dot{X}_i^8)^4=(F^{8}_{0i})^{4}\equiv V^{4}$, one is left 
with the result that the first non-vanishing contribution to the effective
potential has four derivatives,
\begin{equation}
\delta{\cal{L}}_{eff}^{(1)} \sim \frac{V^4}{R^7},
\end{equation}
even when the gauge group experiences multiple levels of breaking.

\subsection{$V^4 x^2/R^9$ and $V^4 v^2/{R^9 r^5}$}

Looking at possible insertions with two background fields
on a massive loop with four derivatives gives terms of the
form,

\begin{equation}
Tr[(\triangle_F\triangle_J)^4\triangle_2']
\label{op}
\end{equation}
\begin{equation}
-\frac{5}{2} Tr[(\triangle_F\triangle_J)^4(\triangle_F\triangle_R)^2]
\label{R}
\end{equation}
\begin{equation}
-\frac{5}{2} Tr[(\triangle_F\triangle_J)^4(\triangle_F\triangle_r)^2]
\label{r}
\end{equation}
\begin{equation}
-5Tr[(\triangle_F\triangle_J)^4(\triangle_F\triangle_R)
(\triangle_F\triangle_r)].
\label{Rr}
\end{equation}
The operators in (\ref{op}) and (\ref{R}) lead to terms
of the form $V^4 x^2/R^9$
with $x$ being a light field (scale $r$) in agreement with \cite{dr} ,
whereas the operators in (\ref{r}) and (\ref{Rr}) give terms with more
powers of $R$ in the denominator.
At this point in our analysis, one might worry that we have thrown out
the vertices coupling three quantum fields (two of mass R and one of mass
r) with one background field which was found to be important in
the result of \cite{fer}.  However, by considering the $x$'s as background
plus quantum fields, the effective operator $V^4 {x'}^2/R^9$ contains the sum
of all non-vanishing vertices with up to four derivatives constructable
from such a vertex.  We can now use $V^4 {x'}^2/R^9$ in the path
integral (\ref{path}) and integrate over the light modes
$x'$ to generate
\begin{equation}
\frac{V^4 v^2}{R^9 r^5},
\label{v^6}
\end{equation}
where $v^2 \equiv (\dot{X}_i^3)^2$.
Clearly (\ref{v^6}) has the wrong dependence on $R$ and
$r$ to reproduce the term of interest in the supergravity scattering amplitude.

\section{Comment on the Eikonal approximation}

When analyzing D0-brane scattering most authors (see e.g. \cite{bb,fer}
and references therein) have chosen to use an explicit background
given by $x=vt+b$ where $v$ is a relative velocity of the
D0-branes and $b$ an impact parameter.  Such an approach allows
one to construct the exact propagator as a power series in $b$, $v$,
and $t$.  By organizing the calculation along the
lines suggested by our analysis above, we can 
%easily 
exhibit the cancellation of all $V^4 v^2/R^7 r^7$
contributions to the effective action.  The point, again,
is to take advantage of the large $R$ limit.  In the functional
integral, one first does the integration over the fields with
mass of order $R$.  As explained in section 2.1 terms involving
only $D^2$ cancel, allowing one to write a simplified expression
for the effective action which only depends on the difference
of the derivative terms between bosons and fermions
\begin{equation}
\Gamma[X]=\frac{i}{2}Trln[1+\triangle_F \triangle_J^B]-
\frac{i}{8}Trln[1+\triangle_F \triangle_J^\psi],
\end{equation}
where $\triangle_F \equiv -D^{-2}$ is the propagator for the
heavy fields and is a function of the background and the light
fields.
Again, terms with two derivatives of the background
or light fields cancel as in (\ref{v2can}).  Terms with four derivatives
and factors of $r^2$ expanded up from the heavy propagator
yield precisely the structure $V^4 x^2/R^9$. So again,
there are no terms of the form $V^4 v^2/{R^7 r^7}$ in the effective action. 

This of course does not mean that there are not individual
diagrams with the behavior $V^4 v^2/{R^7 r^7}$.
However, we see explicitly
from this analysis that there are cancellations between
bosons and fermions.  In \cite{fer}, a particular diagram
with this behavior was exhibited.  But we see that this
contribution is cancelled by diagrams involving fermions.

%Our procedure has the virtue that it does
%not depend on the form of $D^{-2}$, nor how the $SU(3)$ gauge group
%is broken down, it only requires
%the supersymmetric degrees of freedom present in the matrix model to 
%explicitly show the cancellations between bosons and fermions
%leading to our conclusions.

\section{Discussion}

We have shown by explicit calculation in an arbitrary background that 
the operators one first encounters in the matrix model effective
action after integrating out just the most massive modes 
%with any relevance for three graviton scattering 
contain four derivatives
and are of the form
$V^4 x^2/{R^9}$ as discussed in \cite{dr}.
When we use such an operator to construct $V^4 v^2/{R^9 r^5}$ at two 
loops by integrating over the modes of scale r, one finds
the wrong scaling with $R$ and $r$ to correspond with the term
$V^4 v^2/{R^7 r^7}$ in the supergravity $S$-matrix.
In our analysis, we found no velocity independent terms indicating
that fermionic contributions were missed in the work of \cite{fer}.
In fact, in  \cite{wati}\footnote{While this work was being
completed we received word of this result.} the missing fermionic 
piece was identified.
 
What does one conclude about the correspondence between the matrix
model and supergravity for three graviton scattering? It is possible that
a proper treatment of various
subtleties of DLCQ supergravity will
show complete agreement with the finite $N$ matrix model 
\cite{wati}.  It is likely that at large $N$,
the supergravity prediction is recovered. 
The recent
work of \cite{set} showing that there is a non-renormalization
theorem for $v^6$ in $SU(2)$  indicates that the
matrix model-DLCQ supergravity correspondence is working for
the $v^6$ terms, but 
using reasoning similar to \cite{us} it
is not hard to show that some $v^6$ terms are renormalized
in $SU(N)$ for $N \ge 4$.
%as
%shown in \cite{us}, one expects finite
%renormalizations for the
%$v^8$ terms at 2-loops and beyond in $SU(N)$ with $N \ge 3$.
For now, we will have to wait and see
how the issue of three graviton scattering is resolved.
%\newpage

\section{Acknowledgments}
\label{sec-thanks}
We are indebted  to Michael Dine for extremely useful conversations and encouragements, Washington Taylor for
sending us an early preprint of his work with Mark Van Raamsdonk \cite{wati}, and Gautum
Mandal for clarifying some of the aspects of \cite{fer} and \cite{wati}  with us. This work is supported in part by the U.S. Department of Energy. R.E. 
is grateful for GAANN fellowship support.

\noindent
{\bf Note added:} Concurrent with the appearance of this note on hep-th,
the work of \cite{YO} showed conclusively that supergravity
and the matrix model do agree for 3-graviton scattering (including the effect
of recoil \cite{YO2}).
The results reported in this note coincide with these findings since
\cite{YO} have no terms of the form $v_{12}^2 v_{13}^2 v_{23}^2/{R^7 r^7}$
in the effective action of supergravity or the matrix model.  The source
of the error in \cite{dr} occurs in extracting the $S$-matrix from
the matrix model effective action \cite{us2}.


\begin{thebibliography}{99}
\bibitem{bfss} T. Banks, W. Fischler, S.H. Shenker and L. Susskind,
 ``M Theory As A Matrix Model: A Conjecture,'' Phys. Rev. $\textbf{D55}$ (1997) 5112-5128, hep-th/9610043.
\bibitem{sus} L. Susskind, ``Another Conjecture about M(atrix) Theory,'' hep-th/9704080.
\bibitem{sei} N. Seiberg, ``Why is the Matrix Model Correct?,'' Phys. Rev. Lett. $\textbf{79}$ (1997) 3577-3580, hep-th/9710009.
\bibitem{sen} A. Sen, ``D0 Branes on $T^n$ and Matrix Theory,'' hep-th/9709220.
\bibitem{rev}T. Banks, ``Matrix Theory,'' hep-th/9710231; D. Bigatti and L. Susskind, ``Review of Matrix Theory,'' hep-th/9712072; W. Taylor, ``Lectures on D-Branes, Gauge Theory and M(atrices),'' hep-th/9801182; A. Bilal, ``Matrix
Theory: A Pedagogical Introduction,'' hep-th/9710136.
%\bibitem{pol} S. Hellerman and J. Polchinski, ``Compactification in the Lightlike
%Limit,'' hep-th/9711037.
%\bibitem{bil} A. Bilal, ``DLCQ of M-theory as the light-like limit,''
%hep-th/9805070.

%\bibitem{sa} S. Sannan, Phys. Rev. $\textbf{D}$ ???

\bibitem{dr} M. Dine and A. Rajaraman, ``Multigraviton Scattering in the Matrix Model,'' hep-th/9710174.
%\bibitem{od} H. Ooguri, M. Douglas, ``Why Matrix Theory is Hard,''
% hep-th/9710178.
%\bibitem{bc} Berenstein, Corrado, ``M(atrix)-Theory in Various Dimensions,''
% Phys. Lett. $\textbf{B406}$ (1997) 37-43, hep-th/9702108.
\bibitem{fer} M. Fabbrichesi, G. Ferretti and R. Iengo, ``Supergravity and matrix
theory do not disagree on multi-graviton scattering,'' hep-th/9806018.
\bibitem{ps} M. Peskin and D. Schroeder, {\it An Introduction to Quantum Field Theory},
Addison-Wesley (1995).
%\bibitem{ds} M. Dine, N. Seiberg, ``Comments on Higher Derivative Operators in
%Some SUSY Field Theory,`` Phys. Lett. $\textbf{B409}$ (1997) 239-244, hep-th/97%05057.
\bibitem{bb} K. Becker and M. Becker, ``A Two-Loop Test of M(atrix) Theory,'' Nucl. Phys. $\textbf{B506}$ (1997) 48-60, hep-th/9705091.
\bibitem{set} S. Paban, S. Sethi and M. Stern, ``Supersymmetry and Higher Derivative Terms in the Effective Action of Yang-Mills Theories,'' hep-th/9806028.
%\bibitem{bbpt} K. Becker, M. Becker, J. Polchinski and A.A. Tseytlin, 
%``Higher Order Graviton Scattering in M(atrix) Theory,'' Phys. Rev. $\textbf{D%56}$ (1997) 3174-3178, hep-th/9706072.
\bibitem{us}M. Dine, R. Echols and J. Gray, ``Renormalization of Higher Derivative
Operators in the Matrix Model,'' hep-th/9805007.
\bibitem{wati}W. Taylor and  M. Van Raamsdonk, ``Three-graviton scattering in
Matrix theory revisited,'' hep-th/9806066. 
\bibitem{YO} Y. Okawa and T. Yoneya, ``Multibody Intereactions of D-Particles
in Supergravity and Matrix Theory,'' hep-th/9806108.
\bibitem{YO2} Y. Okawa and T. Yoneya, ``Equations of Motion and Galilei
Invariance in D-Particle Dynamics,'' hep-th/9808188.
\bibitem{us2} M. Dine, R. Echols and J. P. Gray, ``Tree Level Supergravity
and the Matrix Model,'' hep-th/9810021.

\end{thebibliography}
\end{document}